# A growth diagram for plasma-assisted molecular beam epitaxy of GaN nanocolumns on Si(111)

S. Fernández-Garrido, J. Grandal, E. Calleja,

M. A. Sánchez-García, and D. López-Romero

*ISOM and Dpto. de Ingeniería Electrónica, ETSI Telecomunicación,*

*Universidad Politécnica de Madrid*

The morphology of GaN samples grown by plasma-assisted molecular beam epitaxy on Si(111) was systematically studied as a function of impinging Ga/N flux ratio and growth temperature (730-850ºC). Two different growth regimes were identified: compact and nanocolumnar. A growth diagram was established as a function of growth parameters, exhibiting the transition between growth regimes, and showing under which growth conditions GaN cannot be grown due to thermal decomposition and Ga desorption. Present results indicate that adatoms diffusion length and the actual Ga/N ratio on the growing surface are key factors to achieve nanocolumnar growth.

* electronic mail: sfernandez@die.upm.es

III-nitride nanocolumns (III-N NCs) are the subject of intense research since the past decade because of their unique properties and potential electronic and opto-electronic applications. NCs are usually grown on Si(111), Si(100), SiC and sapphire substrates by a self-assembly process using plasma-assisted molecular beam epitaxy (PA-MBE) [1-5]. Unlike continuous layers, NCs accommodate the lattice-mismatch with the substrate through a network of misfit dislocations localized at the heterointerface [6]. Therefore, they grow fully relaxed and free of extended defects such as basal plane stacking faults or threading dislocations [6,7]. This fact makes III-N NCs excellent candidates to develop arrays of highly efficient nano-light-emitters in the infrared-visible-ultraviolet range.

Though the very first results on the PA-MBE growth of GaN NCs were reported in the late nineties [1,2], there are still open issues under discussion such as the specific growth conditions that lead to NCs formation, the influence of growth parameters on the NCs morphology, and the driving mechanism of the self-assembly growth process [8-13]. Regarding growth conditions, it is well known that GaN NCs generally grow under N-rich conditions (Ga/N ratio < 1) [2-7,9]. A few groups claimed that GaN NCs could be grown under Ga-rich conditions (Ga/N ratio > 1) at high temperature (~800 ºC) [1,8]. However, the authors neither provided a proper calibration of the impinging fluxes in universal units, nor considered the quantitative effect of the very high Ga desorption rate at 800 ºC [14]. These facts make difficult to determine whether or not the *actual* growth conditions were Ga-rich.

In order to unambiguously establish under which conditions GaN NCs can be grown on Si(111) by PA-MBE, the GaN morphology was systematically studied as a function of impinging Ga/N flux ratio and growth temperature. A growth diagram was built from the results, emphasizing that the NC density, diameter and growth rate depend on these parameters.

All samples were grown on bare Si(111) substrates in a Riber Compact 21s MBE system equipped with a standard Knudsen effusion cell for Ga and an Addon radio-frequency plasma source for active nitrogen. After a standard degreasing, the substrates were outgassed for 30 min at 900 ºC. The growth temperature was measured with an Ircon Modline 3 optical pyrometer calibrated with the 1X1 → 7X7 surface reconstruction transition temperature of Si(111) (860 ºC [15]). Samples morphology was analysed by high resolution scanning electron microscopy (SEM) (CRESTEC CABL-9500C). Cross-sectional SEM of thick Ga-limited GaN films grown at low temperature (680 ºC) provided the Ga flux calibration in GaN growth rate units (nm/min) [14]. Similarly, the active N flux was also calibrated in nm/min using cross-sectional SEM data from N-limited GaN films. A value of 1 nm/min equals 0.064 ML/s, where 1 ML of GaN corresponds to $c/2 = 0.259$ nm or $1.14 \times 10^{15}$ GaN/cm$^2$ areal density.

The dependence of GaN morphology on impinging Ga/N flux ratio and growth temperature was studied in five set of samples grown at different temperatures (730-850 ºC) varying the impinging Ga flux ($\Phi_{Ga}$ = 0.7-9.8 nm/min) while keeping always constant the active N flux ($\Phi_N$ = 6.0 nm/min). Thus, growth changed from N-rich to Ga -rich conditions, in terms of impinging fluxes. The growth time for all samples was 180 min.

The reflection high-energy electron diffraction (RHEED) pattern observed during growth in all cases evolved from streaky, typical of the 7X7 surface reconstruction, to a ringlike one indicating a disorientation of the growing layer. Continuing growth, the ringlike pattern fainted and the one corresponding to a hexagonal structure appeared, as reported by Sonmuang *et al.* [5].

Depending on the specific values of $\Phi_{Ga}$ and growth temperature, the sample morphology consisted in a compact layer or NCs aligned along the (0001) direction [3], as shown in figures 1 (a) and 1 (b), respectively. Figure 1 (c) summarizes the results in a growth diagram where the boundary between compact and nanocolumnar growth regimes is represented as a function of $\Phi_{Ga}$ and growth temperature. The region labelled as *no-growth* indicates the growth conditions for which it was not possible to grow GaN, due to the exponential increase of both Ga desorption and GaN decomposition with growth temperature [14,16]. Notice that higher active N fluxes would lead to a more efficient Ga incorporation, shifting this boundary towards higher growth temperatures [16].

The diagram shows that for a given active N flux, GaN NCs become very difficult to grow below 750 ºC, even for very low $\Phi_{Ga}$ values. In addition, the compact to nanocolumnar boundary shifts towards higher temperatures with increasing $\Phi_{Ga}$ values. The existence of a *threshold temperature* to grow NCs suggests that adatoms diffusion must be high enough to enhance localized nucleation and 3D growth [12]. Otherwise, adatoms will cover the substrate surface homogeneously, giving rise to a compact layer. This is in good agreement with recent results by Landré *et al.* [11] showing that NCs growth can be induced by increasing adatoms diffusion when In is used as surfactant. Regarding the effect of the Ga/N ratio on NCs growth, the results evidence that NCs can be grown under *nominally* N-rich ($\Phi_{Ga}/\Phi_N<1$) or Ga-rich ($\Phi_{Ga}/\Phi_N>1$) conditions. However, in the NC regime, the Ga desorption rate is not negligible [14], and this may result in an *actual* Ga/N ratio lower than one even for $\Phi_{Ga}/\Phi_N>1$. Although most experimental evidences point so far to this situation, specific studies on the *in-situ* measure of the desorbing Ga flux during growth would be necessary to elucidate whether GaN NCs grow exclusively under *actual* N-rich conditions. Finally, we stress that, although the present growth diagram holds quantitatively only for the given $\Phi_N$ value, it gives a clear idea on how to control the NCs growth, even on different substrates.

The samples were studied by plain-view SEM in order to analyse with more detail how morphology depends on growth parameters and which are the mechanisms limiting NCs growth (figure 2). Figures 2 (a) to 2 (c) show the GaN samples morphology evolution with increasing $\Phi_{Ga}$ values at 810 ºC. For the lowest $\Phi_{Ga}$ value (0.7 nm/min) isolated NCs were obtained [figure 2 (a)], while for higher $\Phi_{Ga}$ values, NC density and diameter increased leading eventually to the onset of a compact layer [figure 2 (c)]. Therefore, for a given temperature, the maximum value of $\Phi_{Ga}$ to keep the NC growth regime is limited by NCs merging. A similar trend is observed in Figs 2(d) to (f) for a higher growth temperature (830 ºC). Figures 2(b) and 2(d) illustrate the effect of growth temperature on the samples morphology for a given value of $\Phi_{Ga}$, from high density/diameter NCs to low density/diameter ones. Actually, this morphology evolution is similar to that observed from figures 2 (b) to 2 (a), but using temperature changes instead of $\Phi_{Ga}$ variations. This also points to the rough (but not exact) equivalence of both approximations in terms of *actual* Ga flux considering desorption effects.

Figure 3 shows, quantitatively, the dependence of NC average diameter and NC density on $\Phi_{Ga}$ for two different growth temperatures (810 ºC and 830 ºC). The NC average diameter values were derived from the statistical analysis of about 50 isolated NCs per sample (merged NCs were not considered), and the NC density values were estimated from the analysis of 1 x 1 $\mu m^2$ surface areas (we note that merged NCs were counted as one). As shown in figures 3 (a) and 3 (b), for a given growth temperature, NC diameter steadily increased with $\Phi_{Ga}$; from 18 nm ($\Phi_{Ga}$ = 0.7 nm/min) to 27 nm ($\Phi_{Ga}$ = 4.5 nm/min) and from 21 nm ($\Phi_{Ga}$ = 4.5 nm/min) to 25 nm ($\Phi_{Ga}$ = 6.7 nm/min) for 810 ºC and 830 ºC, respectively. When comparing figures 3 (a) and 3 (b), a clear reduction in NC diameter with growth temperature can be observed. In particular, for $\Phi_{Ga}$ = 4.5 nm/min, it decreased from 27 nm down to 21 nm. Taking into account the raise in NC diameter with $\Phi_{Ga}$, this behavior can be explained considering the exponential increase in Ga desorption [14]. Regarding NC density, at 810 ºC, it rapidly increased with $\Phi_{Ga}$ (from 0.3 x $10^{10}$ NCs/cm$^2$ up to 3.7 x $10^{10}$ NCs/cm$^2$) until NCs merging led to an apparent density reduction (2.5 x $10^{10}$ NCs/cm$^2$) [Fig. 3 (c)]. For 830 ºC, the same trend was observed but shifted towards higher $\Phi_{Ga}$ values due to the higher Ga desorption rate [Fig. 3 (d)].

Figures 4 (a) and 4 (b) summarize the dependence of the growth rate on $\Phi_{Ga}$ and growth temperature respectively, for the given $\Phi_N$ value. The growth rate was derived, from cross-sectional SEM data, as the layer thickness in compact layers and the NC average height divided by the growth time. The NC average height was estimated from the statistical analysis of about 30 NCs per sample. As shown in figure 4 (a) the growth rate was approximately given by the $\Phi_{Ga}$ value in the NC growth regime, while it saturated at a value of 5.2 nm/min within the compact regime, being close to the active N flux value of 6.0 nm/min. This difference may be attributed to GaN losses caused by thermal decomposition [16]. Figure 4 (b) shows the growth rate temperature dependence for $\Phi_{Ga}$ = 4.5 nm/min. Below 810 ºC, the growth rate had a constant value (~ 4.5 nm/min) in good agreement with the nominal one, given by $\Phi_{Ga}$. For temperatures above 810 ºC, within the NC regime, the growth rate steadily decreased due to the exponential increase in both Ga desorption and GaN decomposition [14,16].

In summary, the morphology of GaN samples grown by PA-MBE on Si(111) was systematically studied as a function of impinging Ga/N flux ratio and growth temperature. Two different morphologies were observed: compact and nanocolumnar. A growth diagram was established as a function of growth parameters, exhibiting the boundary between compact and NC growth regimes, and also showing under which conditions it is not possible to grow GaN. Within the NC growth regime: i) NC diameter increases with impinging Ga/N flux ratio and decreases with growth temperature, ii) for a given growth temperature, NC density increases with impinging Ga/N flux ratio until NCs merging occurs, and iii) NC growth rate along the (0001) direction is limited by Ga. Application of this diagram can provide an accurate control of the resulting morphology and a guide to compare the physical properties of GaN samples grown on Si(111).

## Acknowledgments

We acknowledge fruitful discussions with Ž. Gačević and A. Bengoechea. This work was partially funded by the Spanish Ministry Science and Innovation, MICINN Projects MAT2008-04815 and Consolider-CSD 2006-19; and the Community of Madrid, Project CAM S-0505-/ESP-0200.

## References

[1] M. Yoshizawa, A. Kikuchi, M. Mori, N. Fujita, and K. Kishino, Jpn. J. Appl. Phys. **36**, L459 (1997).

[2] M.A. Sánchez-García, E. Calleja, E. Monroy, F.J. Sánchez, F. Calle, E. Muñoz, and R. Beresford, J. Cryst. Growth, **183**, 23 (1998).

[3] E. Calleja, M.A. Sánchez-García, F.J. Sánchez, F. Calle, F.B. Naranjo, E. Muñoz, U. Jahn, and K.H. Ploog, Phys. Rev. B **62**, 16826 (2000).

[4] L. Cerrutti, J. Ristic, S. Fernández-Garrido, E. Calleja, A. Trampert, K.H. Ploog, S. Lazic, and J.M. Calleja, Appl. Phys. Lett. **88**, 213114 (2006).

[5] R. Songmuang, O. Landré, and B. Daudin, Appl. Phys. Lett. **91**, 251902 (2007).

[6] A. Trampert, J. Ristic, U. Jahn, E. Calleja, and K.H. Ploog, Inst. Phys. Conf. Ser. **180**, 167 (2003).

[7] J. Ristic, E. Calleja, M.A. Sánchez-García, J.M. Ulloa, J. Sánchez-Páramo, J.M. Calleja, U. Jahn, A. Trampert, and K.H. Ploog, Phys. Rev. B **68**, 125305 (2003).

[8] Y.S. Park, Seung-Ho Lee, Jae-Eung Oh, Chang-Mo Park, and Tae-Won Kang, J. Cryst. Growth, **282**, 313 (2005).

[9] R. Calarco, R.J. Meijers, R.K. Debnath, T. Stoica, E. Sutter, and H. Lüth, Nano Lett. **7**, 2248 (2007).

[10] R. K. Debnath, R. Meijers, T. Richter, T. Stoica, R. Calarco, and H. Lüth, Appl. Phys. Lett. **90**, 123117 (2007).

[11] O. Landré, R. Songmuang, J. Renard, E. Bellet-Amalric, H. Renevier, and B. Daudin, Appl. Phys. Lett. **93**, 183109 (2008).

[12] J. Ristic, E. Calleja, S. Fernández-Garrido, L. Cerutti, A. Trampert, U. Jahn, and K. H. Ploog, J. Cryst. Growth, **310**, 4035 (2008).

[13] C. T. Foxon, S. V. Novikov, J. L. Hall, R. P. Campion, D. Cherns, I. Griffiths, and S. Khongphetsak, J. Cryst. Growth, **311**, 3423 (2009).

[14] B. Heying, R. Averbeck, L. F. Chen, E. Haus, H. Riechert, and J. S. Speck, J. Appl. Phys. **88**, 1855 (2000).

[15] T. Suzuki and Y. Hirabayashi, Jpn. J. Appl. Phys. **32**, L610 (1993).

[16] S. Fernández-Garrido, G. Koblmüller, E. Calleja, and J. S. Speck, J. Appl. Phys. **104**, 033541 (2008).

**List of Figures**

Figure 1 (a) Tilted SEM image of a GaN compact sample grown with $\Phi_{Ga}$ = 4.5 nm/min and $\Phi_N$ = 6.0 nm/min at 730 ºC. (b) Tilted SEM image of a GaN nanocolumnar sample grown with $\Phi_{Ga}$ = 2.0 nm/min and $\Phi_N$ = 6.0 nm/min at 790 ºC. (c) Growth diagram showing the boundary between compact and nanocolumnar regimes as a function of $\Phi_{Ga}$ and growth temperature for a constant value of $\Phi_N$ (6.0 nm/min). Solid symbols define the growth conditions for the GaN samples of this work.

Figure 2 Plain-view SEM images of GaN samples grown under different conditions. In all cases a $\Phi_N$ of 6.0 nm/min was used.

Figure 3 Nanocolumns diameter (a)/(b) and density (c)/(d) as a function of $\Phi_{Ga}$ for a constant growth temperature (810 ºC/830 ºC) and $\Phi_N$ (6.0 nm/min). NC diameter was derived from the statistical analysis of about 50 isolated NCs per sample and the error bars are given by standard deviation. Solid lines are guides to the eye.

Figure 4 (a) Growth rate at 810 ºC as a function on $\Phi_{Ga}$. (b) Growth rate temperature dependence for $\Phi_{Ga}$ = 4.5 nm/min. In both cases, a $\Phi_N$ of 6.0 nm/min was used. For NC samples, the growth rate was derived from the statistical analysis of the height of about 30 NCs per sample and error bars are given by standard deviation. Solid lines are guides to the eye.

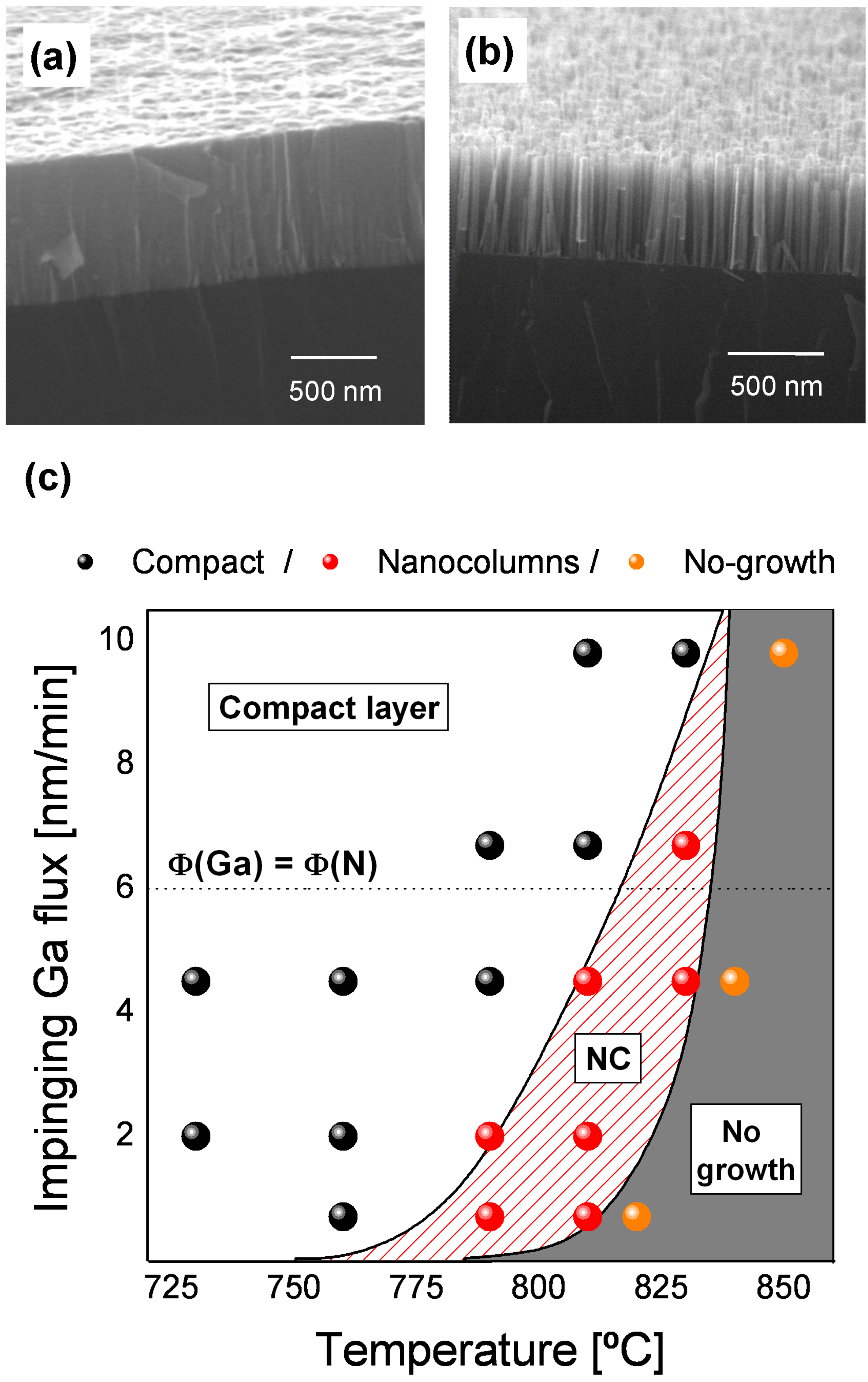

(a)
500 nm
(b)
500 nm
(c)
Compact / Nanocolumns / No-growth
Compact layer
Φ(Ga) = Φ(N)
NC
No growth
Impinging Ga flux [nm/min]
10
8
6
4
2
725
750
775
800
825
850
Temperature [ºC]

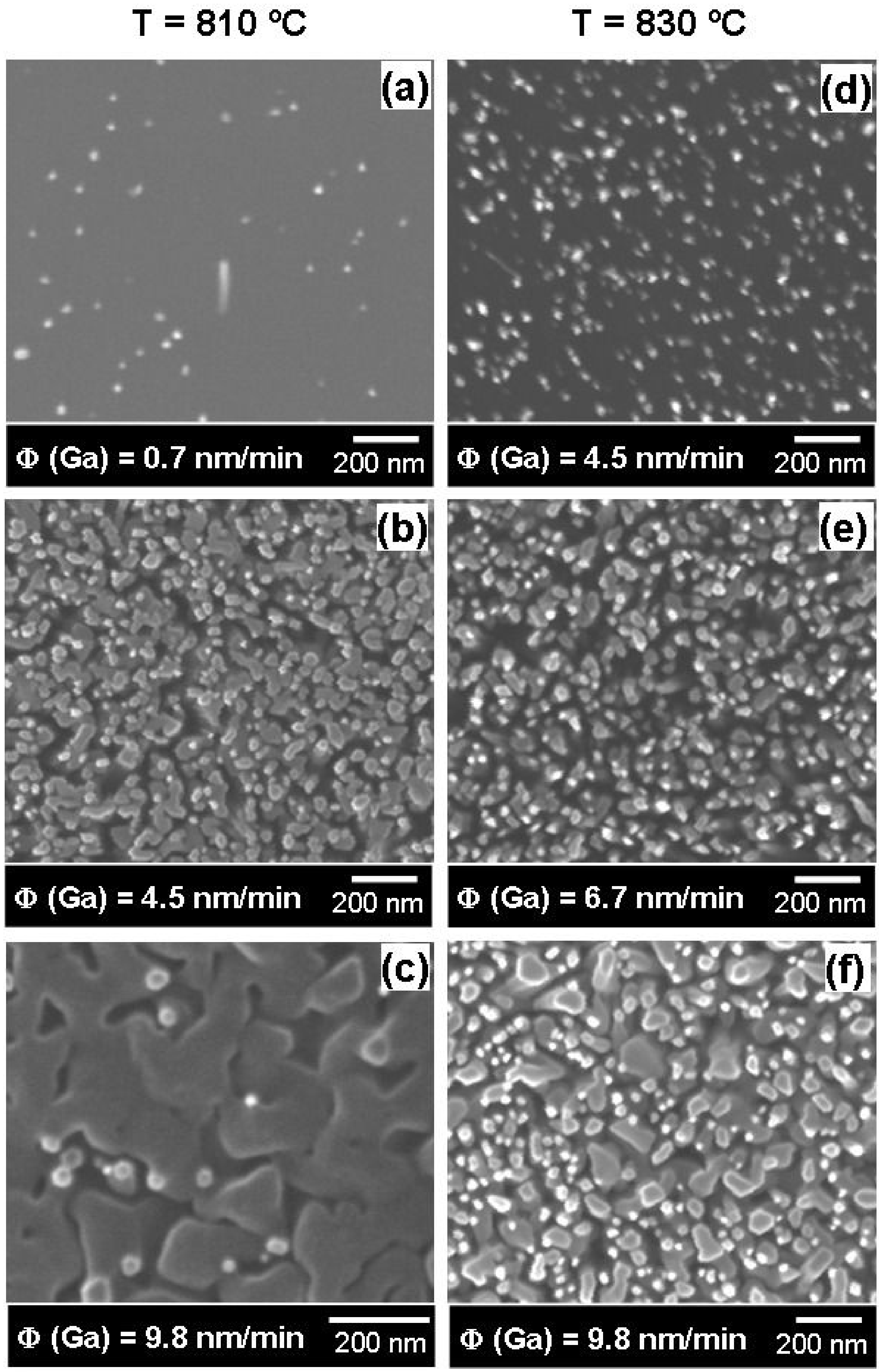
T = 810 ºC
T = 830 ºC
(a)
Φ (Ga) = 0.7 nm/min
200 nm
(b)
Φ (Ga) = 4.5 nm/min
200 nm
(c)
Φ (Ga) = 9.8 nm/min
200 nm
(d)
Φ (Ga) = 4.5 nm/min
200 nm
(e)
Φ (Ga) = 6.7 nm/min
200 nm
(f)
Φ (Ga) = 9.8 nm/min
200 nm

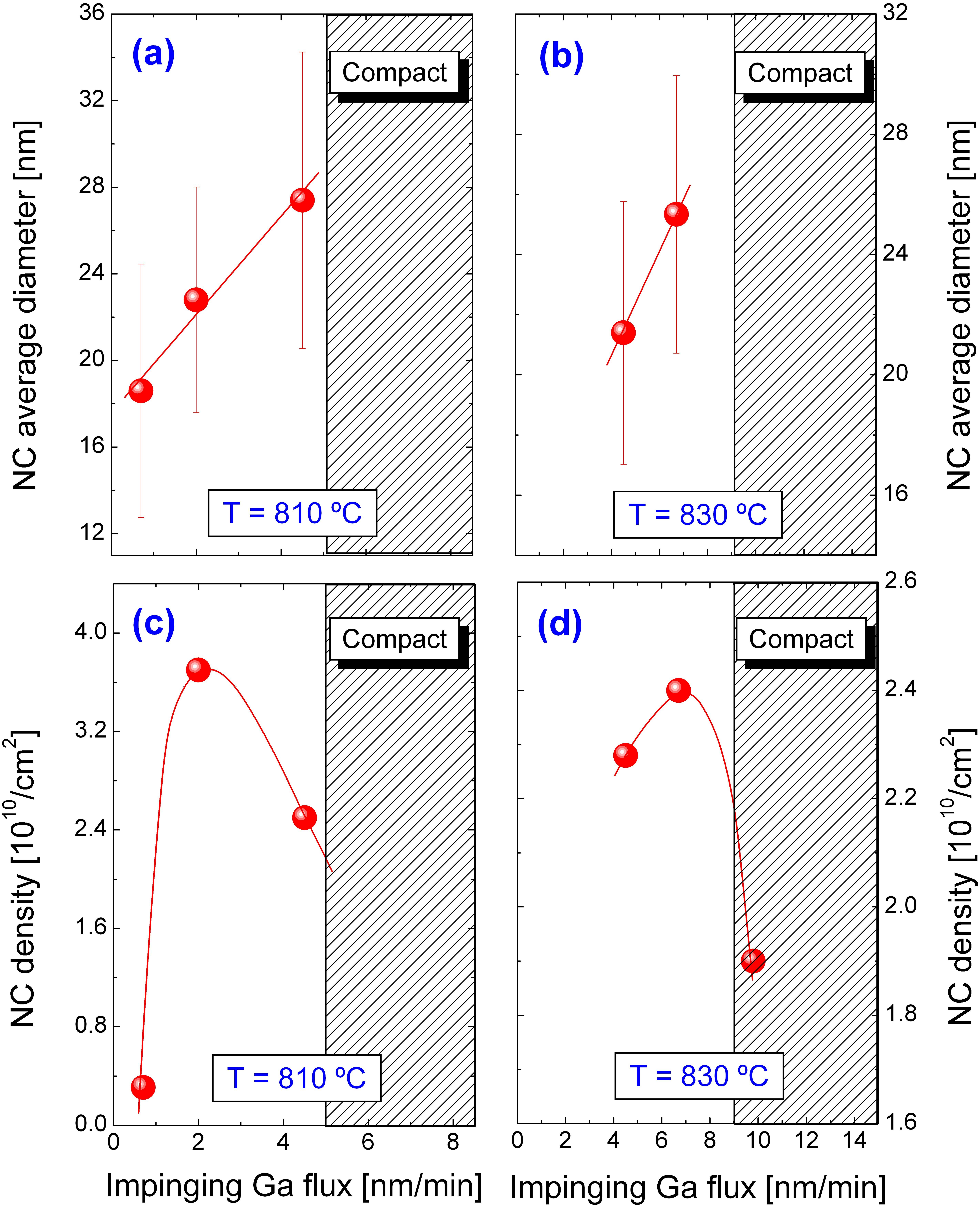

(a)
(b)
(c)
(d)
Compact
T = 810 ºC
T = 830 ºC
NC average diameter [nm]
NC density [10^10/cm^2]
Impinging Ga flux [nm/min]

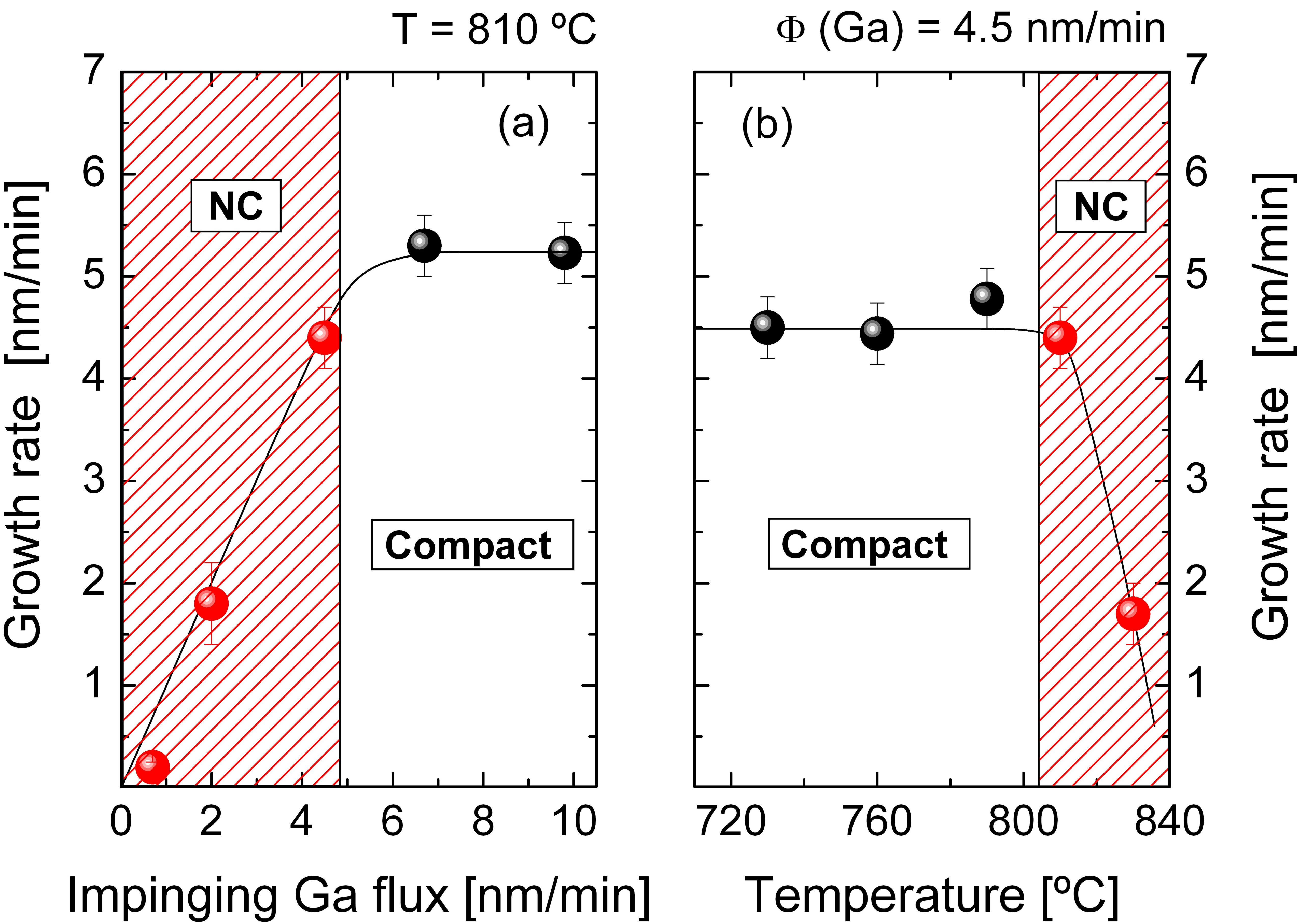

T = 810 ºC
Φ (Ga) = 4.5 nm/min
(a)
(b)
NC
NC
Compact
Compact
Growth rate [nm/min]
Growth rate [nm/min]
Impinging Ga flux [nm/min]
Temperature [ºC]
0
2
4
6
8
10
720
760
800
840
1
2
3
4
5
6
7